\documentclass[twocolumn,prb,floats,superscriptaddress]{revtex4}
\usepackage{graphicx,epsfig}
\usepackage{times}
\usepackage{graphics,dcolumn,bm,float}
\usepackage{amssymb,amsmath,rotate,color}
\usepackage[normalem]{ulem}

\begin{document}

\title{A DFT study on the electronic and magnetic properties of triangular graphene antidot lattices}
\author{Zahra Talebi Esfahani}
\affiliation{Department of Physics, Payame Noor University, P.O.
Box 19395-3697 Tehran, Iran}
\author{Alireza Saffarzadeh}\email{asaffarz@sfu.ca}
\affiliation{Department of Physics, Payame Noor University, P.O.
Box 19395-3697 Tehran, Iran} \affiliation{Department of Physics,
Simon Fraser University, Burnaby, British Columbia, Canada V5A
1S6}
\author{Ahmad Akhound}
\affiliation{Department of Physics, Payame Noor University, P.O.
Box 19395-3697 Tehran, Iran}

\date{\today}

\begin{abstract}
We explore the effect of antidot size on electronic and magnetic
properties of graphene antidot lattices from first-principles
calculations. The spin-polarized density of states, band gap,
formation energy and the total magnetization of two different
equilateral triangular and right triangular antidots with zigzag
and mixed zigzag-armchair edges are studied. We find that although
the values of band gap, formation energy and the total
magnetization of both structures are different, these values may
increase when the number of zigzag edges is increased. The
armchair edges have no contribution in the total magnetization of
right triangular antidots. The induced magnetic moments are mainly
localized on the edge atoms with a maximum value at the center of
each side of the triangles. We show that a spin-dependent band gap
opens up in bilayer graphene as a result of antidot pattern in
only one layer of the structure. Such periodic arrays of
triangular antidots that cause a spin-dependent band gap around
the Fermi energy can be utilized for turning graphene from a
diamagnetic semimetal into a magnetic semiconductor.
\end{abstract}
\pacs{}

\maketitle

\section{Introduction}
Graphene, a single-layer hexagonal lattice of carbon atoms, has
been intensively studied in the past decade due to its remarkable
electronic, optical, mechanical and thermal properties
\cite{Novoselov1,Geim1,Lee1,Neto1}. Despite the intrinsically
positive features of graphene, its gapless band structure and the
lack of intrinsic magnetic property are still a crucial obstacle
in the use of this material as an alternative to silicon in
nanoelectronics. This suggests that turning the semimetallic
graphene into a gapped semiconductor promotes graphene as a key
material in many electronic devices. For instance, band gap
opening in graphene can be used to fabricate transistors with
sufficient on-off ratio \cite{Liao,Wu1}.

To introduce a band gap in graphene, several techniques have been
proposed, including adsorption of suitable elements
\cite{Yavari,Balog}, strain \cite{Guinea}, substrate-induced
symmetry breaking \cite{Zhou}, synthesis of graphene nanoribbons
\cite{Han,Li}, and biased bilayer graphene \cite{Zhang,Castro}.
Moreover, recent studies have shown that a periodic array of holes
(antidots) in graphene is another way to open a band gap or to
demonstrate wave guiding effects in single-layer graphene
\cite{Furst,Vanevic,Peder,Ouyang,Peter,Pedersen1,Power}. The
electronic and magnetic properties of such graphene antidot
lattices (GALs) are strongly dependent on the geometry and edge
orientation. Magnetic edge states induced by creating periodic
arrays of specific antidots in graphene make these magnetic
antidot lattices ideal for spintronic applications. In this
context, a recent experiment \cite{Schneider} has shown that
antidot lattices and their potential for programmability can be
used in magnonic devices.

Furst et al. \cite{Furst} studied the electronic band structure of
a hexagonal lattice with circular holes by means of Dirac
equation, tight-binding calculations and density functional theory
(DFT). It was shown that hydrogen passivation along the edges of
the holes in DFT calculations has a significant influence on the
band structure. Using tight-binding calculations for triangular
antidot arrays, it was reported that edge carbon atoms with
dangling bonds, have higher on-site potentials, even in the case
of hydrogen passivation \cite{Vanevic}. This feature influences
the band structure by shifting and splitting the flat bands around
the Fermi energy \cite{Vanevic}. Pedersen et al. \cite{Peder} have
investigated the hexagonal antidot arrays within the Hubbard model
and found that the energy gap is a factor of $\frac
{\sqrt{N_r}}{N_t}$ where $N_r$ is the number of removed atoms for
creating an antidot, and $N_t$ is the total number of atoms in the
corresponding supercell before antidot formation. The antidot
lattices with hexagonal arrays and zigzag edges have also been
studied in tight-binding methods and DFT calculations by focusing
on bandgap opening/closing \cite{OuyangPCCP} and considering the
effects of inter-antidot distance and the size of antidots on
electronic band structure \cite{Ouyang}.

The previous theoretical works on GALs have mainly focused on
hexagonal and circular holes with particular sizes. Although
theoretical studies on GALs with triangular holes have also been
reported \cite{Vanevic,Liu2009}, these studies which are limited
to equilateral triangular antidots have only been investigated by
single-band tight-binding approximation without considering the
antidot size effects on their electronic properties. Moreover, the
formation energy and the total magnetic moment of these triangular
holes have not yet been reported.

In this paper, we present a study of electronic and magnetic
properties of  triangular-shape graphene antidots, as periodic
hexagonal arrays by means of first-principles calculations. In
order to consider the effects of both the armchair and zigzag
edges of graphene holes on the band gap opening, two different
triangular antidots are examined: Right triangular antidot (RTA)
(see Fig. 1(a)) and equilateral triangular antidot (ETA) (see Fig.
1(b)). We find that the values of band gaps, formation energies,
and local magnetic moments are strongly dependent on the shape and
size of antidots. In addition, we study the influence of antidots
on graphene bilayer structures \cite{Gregersen2015,Kvashnin2015}
composed of a single-layer GAL and a pristine graphene layer
acting as a substrate for the top GAL. Since the edge states are
highly sensitive to the spatial arrangement of the atoms
\cite{Power,Farghadan1} and induce magnetic moments on the
zigzag-shaped edges in graphene nanoribbons and nanorings
\cite{Farghadan2,Fujita,Grujic}, introducing the ETAs with their
pure zigzag edges and also the RTAs with their mixed armchair and
zigzag edges in graphene lattices that we consider here, opens up
spin-dependent band gaps which may provide a guideline for future
magneto-optical experiments in graphene-based nanodevices.
\begin{figure}
\centerline{\includegraphics[width=0.90\linewidth]{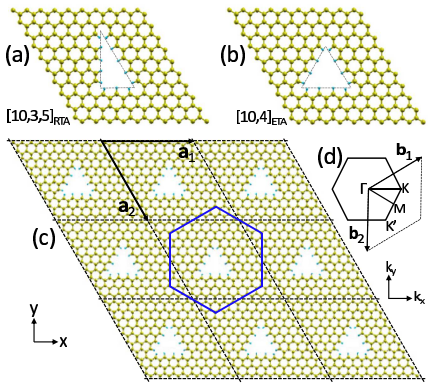}}
\caption{Optimized geometry of $10\times10$ graphene supercells
with (a) $[10,3,5]_{RTA}$ and (b) $[10,4]_{ETA}$. The dashed lines
are guides to the eye. The yellow and blue balls represent carbon
and hydrogen atoms, respectively. (c) Sketch of a $[10,4]_{ETA}$
GAL, obtained from the optimized structure (b). The blue lines
show the corresponding GAL unit cell. The vectors $\textbf{a}_1$
and $\textbf{a}_2$ represent the supercell basis vectors. (d)
Hexagonal (solid) and rhombic (dotted) Brillouin zones of the GAL
with reciprocal vectors $\textbf{b}_1$ and $\textbf{b}_2$.}
\label{image1}
\end{figure}

The paper is organized as follows. The methodology and
computational details are given in Sec. II. In Sec. III, we
present the numerical results of electronic and magnetic
properties of graphene RTA and ETA lattices using DFT
calculations. Band-gap opening in bilayer graphene with antidot
pattern in one of the layers is also discussed. A brief conclusion
is given in Sec. IV.

\section{Computational methods}
To investigate the electronic and magnetic properties of
triangular GALs, we use two notations, $[L, D]_{ETA}$ for ETA and
$[L,Z,A]_{RTA}$ for RTA arrays. Here $L$ denotes the number of
carbon atoms along the supercell edges, $D$ represents the number
of passivated carbon atoms along the ETA sides, and $Z$ and $A$
are the numbers of passivated carbon atoms along the two
perpendicular RTA sides with zigzag and armchair edges,
respectively.

The DFT calculations were performed using the SIESTA code
\cite{sole,orde}, in which the exchange-correlation potential was
approximated by the generalized gradient approximation (GGA)
\cite{pedr}. The Perdew-Burke-Ernzerhof (PBE) exchange-correlation
functional, norm-conserving Troullier-Martins pseudopotentials,
and a double-$\zeta$ polarized basis(DZP) are used for this
calculation \cite{Artacho}. All structures are subjected to
periodic boundary conditions using a $10\times10$ ($L=10$)
supercell geometry with basis vectors $a(1,0,0)$ and
$a(\frac{1}{2},-\frac{\sqrt{3}}{2},0)$ in which $a=24.629$ {\AA},
and vacuum in the $z$ direction for pristine graphene structure.
Figs. 1(a) and (b) show two supercells with $[10,3,5]_{RTA}$ and
$[10,4]_{ETA}$, respectively. The cut off energy is set to 300 eV
and the Brillouin zone sampling is performed by the Monkhorst-pack
mesh of k-points. A mesh of (10$\times$10$\times$1) has been
adopted for discretization of \textbf{k}-points and a broadening
factor of 0.04eV is assumed for the total density of states
(TDOS). Moreover, spin-polarized calculations were performed to
obtain the total magnetic moment. Also, Mulliken population
analysis was used in calculation of local magnetic moment on each
carbon atom.
\begin{figure}
\centerline{\includegraphics[width=0.85\linewidth]{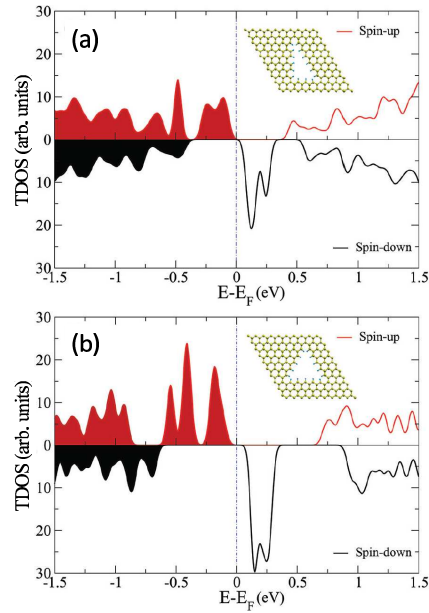}}
\caption{Calculated spin-polarized TDOS for (a) $[10,3,6]_{RTA}$
and (b) $[10,5]_{ETA}$ GALs. The shaded regions indicate the
occupied states with spin-up and spin-down electrons. The insets
show the graphene supercells with corresponding antidots.}
\label{image2}
\end{figure}

To compare the stability between RTA and ETA lattices with
different antidot sizes, the formation energy,
$E_{\mathrm{form}}$, is defined as \cite{Zhang1991,Reich2005}
\begin{equation}\label{1}
E_{\mathrm{form}} = E_{\mathrm{GAL}}-
N_{\mathrm{C}}\mu_{\mathrm{C}} -
\frac{1}{2}N_{\mathrm{H}}E_{\mathrm{H_2}}\ ,
\end{equation}
where $E_{\mathrm{GAL}}$ is the total energy of the GAL supercell,
$\mu_\mathrm{C}$ is the chemical potential of C atom which is
taken as the energy of one carbon atom in the perfect graphene,
and $E_\mathrm{H_2}$ is the total energy of an isolated
$\mathrm{H_2}$ molecule. $N_{\mathrm{C}}$ and $N_{\mathrm{H}}$ are
the number of carbon and hydrogen atoms in the GAL supercell,
respectively.

It is worth mentioning that the formation energy, given in Eq.
(\ref{1}) for GALs, is slightly different with that in the
structures with charged defects \cite{Zhang1991}. In fact, in the
defective structures, $E_{\mathrm{form}}$ depends on the atomic
chemical potentials as well as the electron chemical potential
\cite{Zhang1991}. However, in the GALs consisting of a periodic
array of holes (antidots) with the same shapes and sizes, the
holes differ from charged defects in graphene. Therefore, although
defects can also be deliberately or accidentally introduced into
GALs and affect their energy, in this research we focus only on
defect-free antidot structures passivated by hydrogen atoms
\cite{Vanevic,Furst}.

\begin{figure}
\centerline{\includegraphics[width=0.85\linewidth]{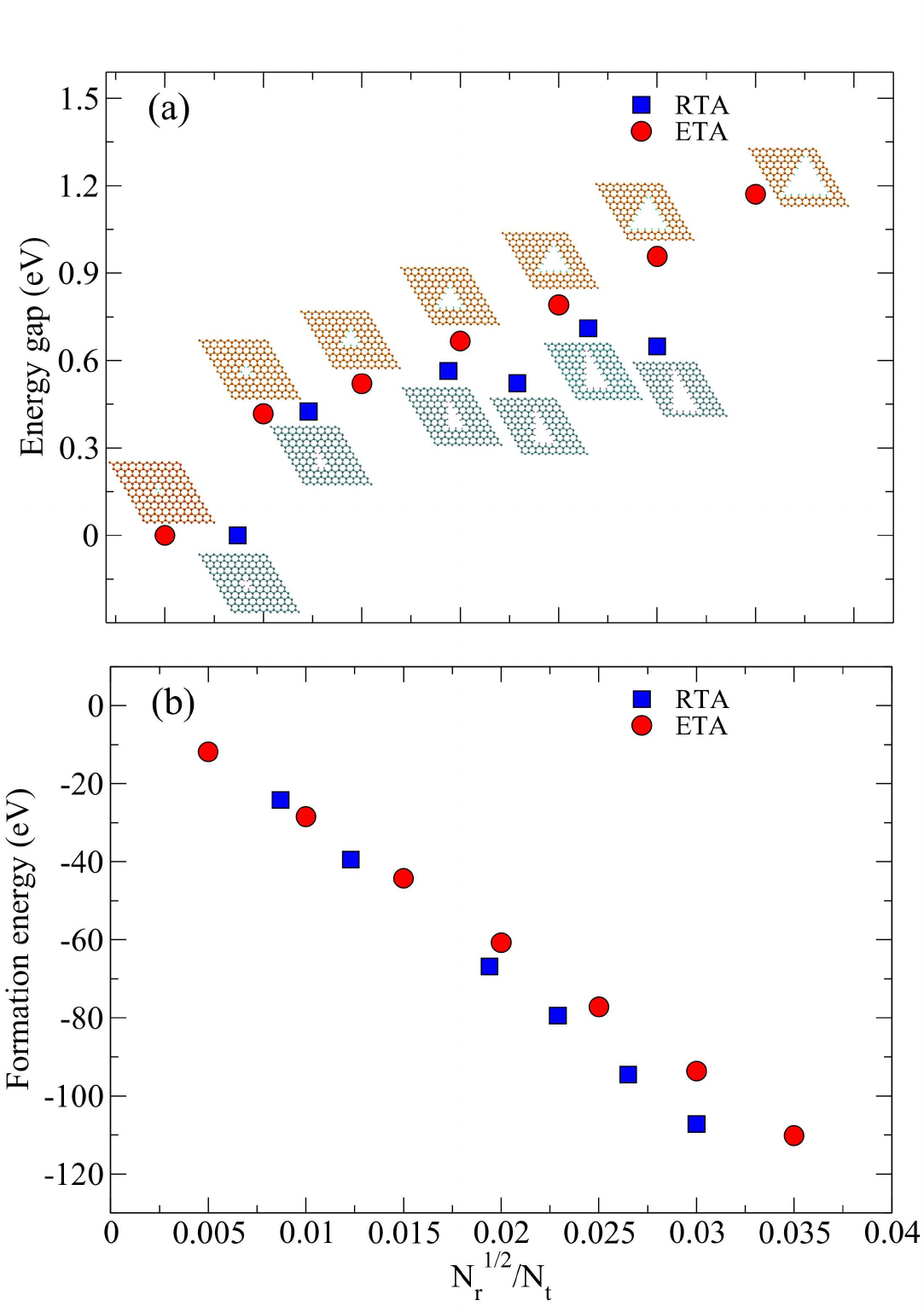}}
\caption{(a) Energy gaps and (b) total formation energies versus
ratio $\frac {\sqrt{N_r}}{N_t}$ for ETA and RTA structures. The
insets in (a) show the corresponding graphene supercells. For
clarity, we have used red and blue colors for carbon atoms in the
ETA and RTA supercells, respectively. In each case, the atoms at
the antidot edges are passivated by hydrogen.} \label{image3}
\end{figure}

\section{results and discussion}
We first investigate how different types of triangular antidots
affect the electronic states. Figures 2(a) and (b) show the
spin-polarized TDOS in $[10,3,6]_{RTA}$ and $[10,5]_{ETA}$
structures, respectively. Comparing the TDOS in the presence of
triangular antidots with that known in pristine graphene clearly
shows the band gap opening and the induced midgap states as a
result of periodic array of antidots. Due to the zigzag-shaped
edges in both structures, localized magnetic moments are induced
on the carbon atoms, which make the antidot lattices appropriate
for device applications. We see that the density of states is
strongly spin dependent with a remarkable difference in their spin
density of states around the Fermi energy. The peaks of TDOS in
the ETA structure are higher in the midgap region due to more
zigzag edges in its supercell. Moreover, the ETA structure shows a
larger band gap compared to the RTA structure, indicating that the
shape of triangular holes has a strong influence on the size of
band gap and the peak of density of states. Interestingly, the
Fermi level is located slightly above the occupied states for
spin-up electrons, while it is slightly below the midgap-states
for spin-down electrons, suggesting a $p$-type ($n$-type)
semiconducting behavior for spin-up (spin-down) electrons in both
triangular antidot structures. This feature demonstrates that the
triangular antidots which exhibit both magnetic and semiconducting
properties, can be utilized for spintronic applications.

We have also studied the band gap and formation energy of GALs
containing triangular antidots with different sizes versus
quantity $\frac{\sqrt{N_r}}{N_t}$ (see Fig. 3). The total number
of atoms in each supercell before creating an antidot is
$N_t$=200. As shown in Fig. 3(a), for the first ETA lattice, i.e.,
[10,1]$_{ETA}$ and also the first RTA lattice, i.e.,
[10,1,2]$_{RTA}$ the band gaps are zero. For the other structures,
however, a gap opens up which roughly increases linearly as both
types of structures are increased in size. This increase is more
obvious in the case of ETA lattices in which the energy gap can
reach 1.17 eV, while it is less than 0.65 eV for the case of RTA
lattices which show small deviations from gap increase as a result
of single dangling bonds in the triangle side with armchair edges.
We see that the value of band gap depends on the size and the
shape of antidots. In other words, antidot-antidot interaction
which has a profound impact on the band gap is directly related to
the antidot density, i.e., the antidot separation. Since this
interaction and the overlap between antidots are included in our
\textit{ab initio} calculations, the obtained results can present
an accurate description of energy gap values compared to the
reported results using tight-binding methods \cite{Gunst} or Dirac
Hamiltonian \cite{Peder}. The dependence of band gap opening on
the size and the shape of antidots suggests the GALs as
semiconducting materials with tunable energy gaps.
\begin{figure}
\centering\includegraphics[width=0.45\textwidth]{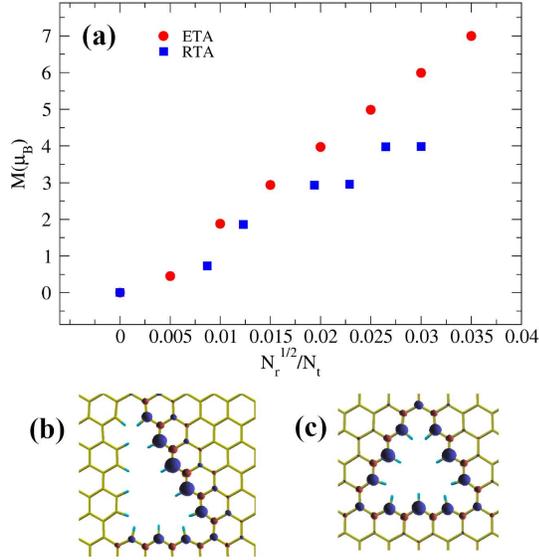}
\caption{(a) Total magnetic moments versus
$\frac{\sqrt{N_r}}{N_t}$ in ETA and RTA structures. Localized
magnetic moments in (b) [10,3,5]$_{RTA}$ and (c) [10,3]$_{ETA}$
lattices. The blue and red regions around carbon atoms in (b) and
(c) represent the localized magnetic moments, while their sizes
denote the magnitude of the moments.} \label{image4}
\end{figure}

The total formation energies for different sizes of ETA and RTA
antidots are depicted in Fig. 3(b). We see that the formation
energy decreases linearly as the number of removed carbon atoms
and the added hydrogen atoms is increased. This feature is
independent of the antidot shape, indicating that the spatial
arrangement of atoms at the edges has a negligible effect on
$E_\mathrm{form}$. As the size of antidots is increased, creating
a RTA lattice becomes slightly more energetically favorable than
creating an ETA lattice, while the energy gaps in ETA structures
are slightly larger than those in RTA structures. Interestingly,
the linear reduction of formation energy by increasing the size of
similar ETA lattices but passivated by oxygen atoms has also been
reported in single-layer MoS$_2$ antidots using DFT calculations
combined with experiments \cite{Gan2016}. Moreover, it is
suggested that since the formation energy of all antidot sizes and
shapes is negative, it is probably feasible to create them by
heating the samples in air \cite{Gan2016}. Note that the value and
the sign of formation energy of unpassivated defective graphene
\cite{Thrower,Haldar}, which cause lattice distortion around
defects, can significantly change when the edges are terminated by
hydrogen atoms. In other words, structural stability of GALs can
be achieved by passivating carbon dangling bonds at the edges.

As we mentioned above, the triangular antidots induce magnetic
moments at the edges and their neighboring carbon atoms, and
hence, the density of states becomes spin dependent. The induced
magnetic moments can change with varying the shape and also the
size of antidots. In Fig. 4(a), the total magnetic moments,
\textbf{M}$=\sum_{i=1}^{n} (\mathbf{m}_{i}^{\uparrow} -
\mathbf{m}_{i}^{\downarrow})$, are depicted versus ratio
$\frac{\sqrt{N_r}}{N_t}$ for ETA and RTA lattices. Here,
$\mathbf{m}_i^{\uparrow(\downarrow)}$ represents the magnetic
moment of spin-up (spin-down) electrons, localized at site $i$ in
the supercell. In both structures the total magnetic moment
increases with increasing the antidot size. Interestingly, the
increase in \textbf{M} is linear in the ETA structures due to the
perfect zigzag edges in each side of the triangle, while the RTA
structures, which consist of mixed armchair and zigzag edges,
exhibit smaller values of magnetic moment compared to those in ETA
supercells. Moreover, in most cases the total magnetic moment
between two successive values of $\frac{\sqrt{N_r}}{N_t}$
increases by $\sim \mu_B$, due to the saturation of dangling bonds
by hydrogen passivation.

We note that in the tight-binding description for a bipartite
lattice structure, such as graphene with two sublattices A and B,
and only nearest-neighbor coupling, the total magnetic moment per
supercell can be determined from Lieb's theorem
\cite{Lieb,Palacios} using the relation
\textbf{M}=$\mu_B|N_A-N_B|$, where $N_{A(B)}$ is the number of
atoms in sublattice A(B) in the supercell. Indeed, the theorem
permits to predict the magnetic moment of the ground state by
simple counting of the sublattice imbalance. Nevertheless, this
theorem does not provide information about the actual local
magnetic order or spin texture. For instance, the local magnetic
moments may couple antiferromagnetically, while \textbf{M}=0. In
contrast, our DFT calculations go beyond the first-neighbor
hopping and intersite Coulomb repulsion (Hubbard model) on which
the theorem is based.
\begin{figure}
\centerline{\includegraphics[width=0.98\linewidth]{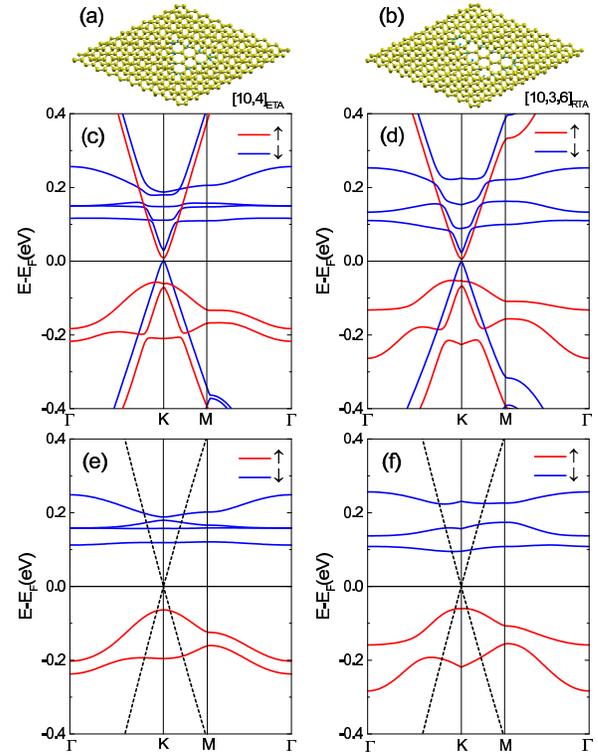}}
\caption{Optimized geometry of $10\times10$ bilayer graphene
supercells with (a) $[10,4]_{ETA}$ and (b) $[10,3,6]_{RTA}$. The
yellow and blue balls represent carbon and hydrogen atoms,
respectively. Spin-dependent energy bands of (c,d) bilayer and
(e,f) single layer graphene in $[10,4]_{ETA}$ and $[10,3,6]_{RTA}$
lattices. The energy bands of pristine graphene are shown by
dashed lines.} \label{image5}
\end{figure}

To compare the distribution of magnetic moments in the two
structures, we have shown in Figs. 4(b) and (c) the localized
magnetic moments on carbon atoms in [10,3,5]$_{RTA}$ and
[10,3]$_{ETA}$ structures, respectively. The blue and red regions
represent opposite directions for the moments. Also, the size of
each region indicates the magnitude of each moment. As expected,
no magnetic moments are formed on the armchair edges, whereas
remarkable moments in both structures can be clearly seen on all
zigzag edges \cite{Saffar-Fargh}. The magnetic moments in the same
sublattices couple ferromagnetically, while the magnetic moments
in different sublattices couple antiferromagnetically.
Nevertheless, the total magnetization in such antidots is not zero
due to the difference in the total magnetization of A and B
sublattices \cite{Farghadan3,Fern}. Moreover, the difference
between the distributions of magnetic moments in RTA and ETA is
remarkable. We find that the total magnetic moment of RTA lattices
does not change, unless the number of armchair edges is increased
with increasing the antidot size. On the other hand, the total
magnetic moment of ETA lattices always increases as the zigzag
edges are increased (see Fig. 4(a)).

Finally, we consider a bilayer structure consisting of a layer of
GAL on top of a pristine graphene layer, arranged in AB stacking.
Here, the pristine graphene can act as a substrate for the GAL.
Such structures can be realized experimentally, by either standard
lithography \cite{Oberhuber,Bai} or vdW stacking techniques
\cite{Geim}. The optimized geometries and the band structures of
[10,4]$_{ETA}$ and [10,3,6]$_{RTA}$ bilayer GALs are depicted in
Figs. 5(a,c), and 5(b,d), respectively . The optimized distance
between the top and bottom layers in both ETA and RTA lattices was
found to be 3.298 {\AA}, which is very close to the interlayer
spacing 3.321 {\AA} in pristine bilayer graphene, computed by our
DFT calculations. From Figs. 5(c) and (d), it is evident that the
introduction of triangular antidots in the bilayer graphene opens
up spin-dependent band gaps, $E_{G\sigma}$, which are much smaller
than those of the single-layer GALs. The band gaps for both
spin-up and spin-down electrons are the same in single-layer GALs,
while in the bilayer GALs, they are $E_{G\uparrow}=68$ meV and
$E_{G\downarrow}=30$ meV in [10,4]$_{ETA}$ lattice and
$E_{G\uparrow}=60$ meV and $E_{G\downarrow}=22$ meV in
[10,3,6]$_{RTA}$ lattice. The main reason for such smallness in
band gaps is due to the creation of antidot pattern in only the
top layer of bilayer graphene. Nevertheless, as we have shown in
Fig. 3(a) for single-layer GALs, the energy gaps of bilayer GALs
are also tunable by changing the antidot size.

To understand the influence of coupling between the top layer with
antidots and the bottom layer without antidots on the
spin-resolved energy bands of bilayer GALs, we have shown in Figs.
5(e) and 5(f) the distinct energy bands of single-layer GALs and
the pristine graphene. We see that in the case of single-layer
GALs and within the energy window around the Fermi level, the
spin-up (spin-down) energy bands are induced only in the lower
(upper) half of the energy window, whereas in the case of bilayer
GALs the spin-up and spin-down bands are formed in both sides of
the Fermi level as a result of the coupling between single-layer
GAL and the pristine graphene in the supercell. By comparing the
energy bands of Figs. 5(c,d) with those of Figs. 5(e,f), we find
that the interaction between the two layers and also between the
magnetic edge states in the top layer can strongly modify the spin
subbands and the energy gaps in the bilayer GALs.

Our calculations for the total magnetic moment of bilayer graphene
result $\textbf{M}=3.968 \mu_B$ for the ETA structure, while we
obtain $\textbf{M}=2.948 \mu_B$ for the RTA structure. These
values are almost the same as the corresponding values in
single-layer GALs (see Fig. 4(a)), suggesting that the interlayer
couplings between the two layers in bilayer GALs mostly affect the
energy gaps. Furthermore, we see that the Fermi level lies at the
minimum of the spin-up conduction band, while it is located at the
maximum of the spin-down valence band. In analogy with the
single-layer GALs, this behavior reveals that the bilayer GALs can
act as $n$-type ($p$-type) semiconductors for spin-up (spin-down)
carriers. Moreover, the bilayer GALs can exhibit the band gap
values which cannot be produced by single-layer GALs. Therefore,
one can selectively tune the energy gap and control the spin of
charge carriers either by single-layer GALs or by bilayer GALs.

It is well-known that single-layer graphene has large mobility due
to the massless Dirac electrons with linear dispersion
\cite{Schwierz}. On the other hand, it has been demonstrated that
the low-energy properties of AB bilayer graphene has two parabolic
bands touching each other at the Fermi energy \cite{Rozhkov}. This
feature implies an effective mass for charge carriers and hence a
lower mobility which may reduce the device performance of bilayer
graphene \cite{Schwierz}. Therefore, since the gap opening in the
single-layer GALs is accompanied by a reduction of charge
mobility, the introduction of periodic holes in only one single
layer of bilayer graphene can create linear dispersion and enhance
the mobility of charge carriers, while the system is still gapped.
We note that the result of DFT calculations of a quantum system
depends on the type of exchange-correlation functional used. For
example, a hybrid functional may overestimate band gaps for
semiconductors but underestimate them for insulators
\cite{Kummel}. In contrast, many-body methods based on the use of
Green's functions and perturbation theory as the GW approximation
can give more accurate predictions and may change quantitatively
the energy gap of GALs. Therefore, a separate numerical
investigation of electronic properties of GALs using
exchange-correlation potential within the GW approximation is also
of practical importance.

\section{conclusions}
Based on the first-principles simulations we have explored the
electronic and magnetic properties of equilateral triangular and
right triangular arrays of antidots with different sizes in
graphene lattices. We have shown that a spin-dependent band gap is
induced in these structures, so that the size of the band gap can
be tuned by varying the antidot size and also the inter-antidot
separation. The shift of Fermi energy towards valence (conduction)
band predicts a $p$-type ($n$-type) semiconducting behavior for
spin-polarized electrons in graphene ETA and RTA lattices. The
local magnetic moments at armchair edges are zero indicating that
these edges do not contribute in the total magnetization of the
system. The variation of antidot size in both RTA and ETA lattices
shows that the energy gap, formation energy, and the total
magnetization increase as the size of antidots is increased.

Moreover, the introduction of an antidot pattern in only one
single layer of bilayer graphene opens up a band gap while the
energy dispersion remains almost linear around the Fermi energy.
Our findings show that the triangular graphene antidot lattices
can be utilized as a platform for creating magnetic semiconducting
materials with tunable band gaps by varying the size of antidots.

\section*{Acknowledgement}
We would like to acknowledge supercomputer time provided by the
WestGrid and Compute Canada.

\end{document}